\documentclass[12pt]{article}
\usepackage{amsmath}
\usepackage[dvips]{graphicx}
\usepackage{epsfig}
\usepackage{amsmath}
\usepackage{amssymb}
\usepackage{graphics,epstopdf}
\usepackage{amsfonts}
\usepackage{amsthm}
\usepackage[usenames]{color}

\definecolor{AV}{rgb}{0.65,0.0,0}
\definecolor{GC}{rgb}{0,0.0,0.65}
\definecolor{WS}{rgb}{0,0.65,0}

\usepackage[
      colorlinks=true,
      linkcolor=blue,
      urlcolor=blue,
      filecolor=blue,
      citecolor=blue,
      pdfstartview=FitV,
      pdftitle={},
      pdfauthor={},
      pdfsubject={},
      pdfkeywords={},
      pdfpagemode=None,
      bookmarksopen=true
]{hyperref}

\newcommand{\bm}{\begin{multiline}}
\newcommand{\beq}{\begin{equation}}
\newcommand{\eeq}{\end{equation}}
\newcommand{\beqs}{\begin{eqnarray}}
\newcommand{\eeqs}{\end{eqnarray}}

\newcommand{\ra}{\rightarrow}

\begin{document}

\thispagestyle{empty}

\hfill{}

\hfill{}

\hfill{}

\vspace{32pt}

\begin{center}

\textbf{\Large Charging axially symmetric interior solutions in General Relativity}

\vspace{48pt}

\textbf{Cristian Stelea,}\footnote{Corresponding author e-mail: \texttt{cristian.stelea@uaic.ro}}
\textbf{Marina-Aura Dariescu,}\footnote{E-mail: \texttt{marina@uaic.ro}}
\textbf{Ciprian Dariescu, }\footnote{E-mail: \texttt{ciprian.dariescu@uaic.ro}}

\vspace*{0.2cm}

\textit{$^1$ Department of Exact and Natural Sciences, Institute of Interdisciplinary Research,}\\[0pt]
\textit{``Alexandru Ioan Cuza" University of Iasi}\\[0pt]
\textit{11 Bd. Carol I, Iasi, 700506, Romania}\\[.5em]

\textit{$^{2,3}$ Faculty of Physics, ``Alexandru Ioan Cuza" University of Iasi}\\[0pt]
\textit{11 Bd. Carol I, Iasi, 700506, Romania}\\[.5em]

\end{center}

\vspace{30pt}

\begin{abstract}
We present two solution-generating techniques, which are direct generalizations of certain Ehlers-Harrison transformations in the Ernst formalism, while adapted to work in presence of an anisotropic fluid source with axial symmetry. Based on these procedures, we were able to construct the electrically charged and the magnetized solution for any static axially-symmetric geometry, which is sourced in general by an anisotropic fluid described by a non-diagonal anisotropic stress-energy tensor. As our main examples we derived and analyzed two new exact solutions with axial symmetry that describe the electrically charged Zipoy-Vorhees interior solution as well as the magnetized Zipoy-Vorhees interior solution, and presented some of their properties. As further examples of our solution-generating techniques we show how to derive two new solutions describing the electrically charged version of the Bowers and Liang solution, as well as a magnetized version of an exact solution with axial symmetry. 
\end{abstract}

\vspace{32pt}

\setcounter{footnote}{0}

\newpage

\section{Introduction}

The search for exact and physically relevant solutions of the Einstein equations has been a topic of renewed interest ever since the beginning of General Relativity (GR) in $1915$. While solving Einstein's equations in their general form presents itself as a formidable task, being a nonlinear system of second order partial derivatives of the metric, during the last century there has been developed an equally impressive arsenal of methods and special techniques to solve them. There is by now an exhaustive collection of exact solutions of Einstein's equations in four dimensions, as collected in \cite{Stephani}.

Moreover, on the experimental side, with the advent of gravitational-wave astronomy \cite{LIGOScientific:2016aoc}, \cite{LIGOScientific:2020ibl} and of the very long baseline interferometry \cite{EventHorizonTelescope:2019dse} one is now able to glimpse into a new physics of the compact objects and black holes that was until now beyond our experimental reach (see for instance \cite{Barack:2018yly} and the references within). Undoubtedly, the black holes play a central role in this field and their properties have been studied extensively (albeit theoretically at first). One of their characteristic features is their uniqueness, as it was best enunciated in Wheeler's famous "black holes have no hair" statement. Basically this means that all (four-dimensional) electrovacuum black-hole spacetimes are characterized by their mass, angular momentum and electric charge, or, in other words, they belong to the class of Kerr-Newman black holes (for a recent  review of the no-hair theorems see \cite{Chrusciel:2012jk} and references therein). 

On the other hand, the physics of the compact objects in GR is equally important to understand the nature of other compact objects which are present in our Universe, such as nuclear stars, white dwarfs, exotic stars, etc. Since the pioneering work of Schwarzschild \cite{Schwarzschild:1916uq} and Tolman \cite{Tolman:1939jz}, compact objects were usually modeled in GR by using spherically symmetric perfect fluid solutions of the Einstein field equations. This was the natural starting point since static perfect fluid configurations seem to lead to spherically symmetric configurations as well \cite{masoud}. As such, in the last century there has been a lot of interest in generating new exact solutions describing spherically symmetric relativistic stars sourced by perfect fluids. There are known by now various procedures to generate new spherically symmetric exact solutions, which are sourced by perfect fluids \cite{Lake:2002bq}, \cite{Martin:2003jc}, \cite{Boonserm:2005ni}. However,  there is a caveat which comes with this plethora of new solutions: as shown in \cite{Delgaty:1998uy}, not all the perfect fluid interior solutions generated this way are physical. Most of the known solutions fail some physical tests hence only some of them seem appropriate to describe isotropic compact objects in GR. This basically means that not every solution generated by these algorithms satisfies the physical requirements and that one should check every generated solution on a case by case basis. 

On the other hand, in modeling realistic compact objects in GR one should consider more sophisticated models, involving deviations either from the spherical symmetry and/or the perfect fluid distribution of the source. Dropping the perfect fluid requirement, while still preserving the requirement of spherical symmetry, one can consider anisotropic fluids  as sources (see for instance \cite{Harko:2002db} - \cite{Herrera:2007kz} and references therein). For such fluids the radial pressure component $p_r$ is not equal to the components in the transverse directions, $p_t$. There are strong theoretical reasons to believe that in realistic stellar models in the high density regimes the pressures inside the star are anisotropic \cite{Ruderman:1972aj}.  Such anisotropies in the fluid distributions can arise from various reasons: they can be due to a mixture of two fluid components \cite{Letelier:1980mxb}, elasticity of the compact objects \cite{Alho:2021sli,Karlovini:2002fc},  the existence of a superfluid phase, the presence of a magnetic field, etc. (for a review see \cite{Herrera:1997plx} and references there). For example, anisotropic fluid models of neutron stars could be used to model the so-called magnetars \cite{DT}, which denote a class of neutron stars whose emissions are powered by the decay of their huge magnetic field. For a magnetar the magnetic field strength can reach values as high as $10^{11}T$, while being even more intense inside the star. There are over $30$ magnetars catalogued by now \cite{Olausen:2013bpa} \footnote{See also the website http://www.physics.mcgill.ca/~pulsar/magnetar/main.html.}(for recent reviews of their properties see \cite{Esposito:2018gvp}, also \cite{Woods:2004kb}). This class of objects includes the soft gamma repeaters (SGRs) and the the anomalous X-ray pulsars (AXPs). Analytic non-perturbative solutions in GR describing anisotropic models of magnetars have been constructed in \cite{Yazadjiev:2011ks} and \cite{Stelea:2018cgm} starting from spherically symmetric solutions. In those works it was shown that in presence of very strong magnetic fields one is forced to consider an axially symmetric treatment of the source (see also \cite{Negreiros:2018cjk}). Moreover, since the astrophysical formation processes of nuclear stars are actually asymmetric in nature, the spherical symmetry is in fact an idealization and one should start from the beginning with axially symmetric interior models of nuclear stars \cite{Raposo:2020yjy}.

One can reach the same conclusion about the necessity of an axially symmetric ansatz to model a realistic compact object if one takes into account its rotation: the angular momentum of the compact object will define a preferred direction and, for stationary configurations, one can assume that the direction of the angular momentum defines the symmetry axis. This is what happens in the celebrated Kerr solution of the vacuum Einstein field equations (see for instance \cite{Chandrasekhar}).
However, even in absence of rotation, the most general line element describing such relativistic systems with axial symmetry has the form \cite{Herrera:2013hm}:
\beqs
ds^2&=&-A(r,\theta)^2dt^2+B(r,\theta)^2dr^2+C(r,\theta)^2d\theta^2+D(r,\theta)^2d\varphi^2.
\label{initialm}
\eeqs
In general, this line element can be considered as a solution of Einstein's field equations\footnote{Note that we work using the natural units for which $G=c=1$.} $G_{\mu\nu}=8\pi T_{\mu\nu}$ sourced by an anisotropic fluid, which is described by a non-diagonal stress-energy tensor of the form \cite{Hernandez-Pastora:2016ctg}:
\beqs
T_{\mu\nu}&=&\rho u_{\mu}u_{\nu}+p_r\chi_{\mu}\chi_{\nu}+p_{\theta}\xi_{\mu}\xi_{\nu}+p_{\varphi}\zeta_{\mu}\zeta_{\nu}+2p_{r\theta}\chi_{(\mu}\xi_{\nu)},
\label{initialf1}
\eeqs
where $\rho$ is the fluid energy density, $p_r$ is the radial pressure, while $p_{\theta}$, $p_{\varphi}$ and $p_{r\theta}$ are transverse components of the fluid pressure. Also $u_{\mu}=(-A, 0, 0, 0)$ is the $4$-velocity of the fluid, while $\chi_{\mu}=(0, B, 0, 0)$, $\xi_{\mu}=(0, 0, C, 0)$ and $\zeta_{\mu}=(0, 0, 0, D)$ are spacelike unit vectors in the radial and transverse directions. 

Of course, not every geometry (\ref{initialm}) leads to physical compact objects describing physically reasonable fluid source configurations. At a bare minimum the energy conditions have to be satisfied by this fluid (see for instance \cite{Maeda:2018hqu}-\cite{Martin-Moruno:2021niw} and references therein). There is also the problem of matching these interior fluid configurations to exterior vacuum geometries, which are vacuum solutions of Einstein's field equations \cite{Herrera:2013hm}. In absence of rotation, for a spherically symmetric compact object, the Birkhoff theorem assures us that the only vacuum solution is (at least part of) the Schwarzschild solution.  Moreover, any spherically symmetric and asymptotically flat solution of the Einstein-Maxwell field equations must be static, so that the exterior geometry of a spherically symmetric charged star must be given by the Reissner–Nordström black hole solution. However, the situation becomes more complicated for axisymmetric sources. For instance, the vacuum region outside a spinning compact object is not generically described by the Kerr geometry \cite{Raposo:2018xkf}. One can understand this in light of the black hole uniqueness and the no-hair theorems: for  the Kerr black hole the Geroch-Hansen multipole moments \cite{Geroch,Hansen} take a very specific form, being determined by the mass and the angular momentum values. One would expect then that the multipole moments of a general spinning compact object might have a more general configuration than that of a Kerr black hole \cite{Frutos-Alfaro:2016arb}. Therefore, the exterior geometry of a stationary axisymmetric fluid configuration should correspond to a vacuum stationary and axisymmetric solution of the Einstein field equations, generically different from the Kerr geometry. 

For static configurations the exterior geometry must then belong to the so-called Weyl-Papapetrou class of axisymmetric metrics \cite{Weyl:1917gp}:
\begin{eqnarray}\label{weyl}
ds_4^2&=&-e^{-\psi}dt^2 +e^{\psi}\big[e^{2\mu}(d\rho^2+dz^2)+\rho^2d\varphi^2\big].
\end{eqnarray}
The metric is specified by the values of two functions $\psi$ and $\mu$, which are functions of the canonical Weyl variables $\rho$ and $z$. For a vacuum solution $\psi$ is a harmonic function in the usual $3$-dimensional Euclidean space with metric $d\rho^2+dz^2+\rho^2d\varphi^2$. Once one knows $\psi$ then the remaining function $\mu(\rho,z)$ is found by performing a simple line-integral using the relations: 
\begin{eqnarray}\label{gammap1}
\partial_z{\mu}=\frac{\rho}{2}\partial_{\rho} \psi\partial_z \psi,~~~~~~~\partial_{\rho}{\mu}=
\frac{\rho}{4}\left[\left(\partial_{\rho} \psi\right)^2-\left(\partial_{z} \psi\right)^2\right].
\end{eqnarray}
The Einstein field equations in four dimensions for a static axisymmetric background are now essentially reduced to finding a solution of Laplace's equation on flat space. Then, every static axisymmetric interior solution (\ref{initialm}) and (\ref{initialf1}) should be matched to a static vacuum solution belonging to the Weyl class (\ref{weyl}). As an example, one interesting vacuum solution with a generic quadrupole moment is given by the so-called Zipoy-Vorhees solution \cite{Zipoy}:
\beqs
ds^2&=&e^{-\psi}dt^2+e^{\psi}\big[e^{2\mu}(d\rho^2+dz^2)+\rho^2d\varphi^2\big],\nonumber\\
e^{-\psi}&=&\left(1-\frac{2m}{r}\right)^{\gamma},~~~~e^{2\mu}=\left(\frac{r(r-2m)}{(r-m)^2-m^2\cos^2\theta}\right)^{\gamma^2},
\label{Zipoy}
\eeqs
where $\rho=\sqrt{r(r-2m)}\sin\theta$ and $z=(r-m)\cos\theta$ are the canonical Weyl coordinates. For integer values of $\gamma$ this metric describes the superposition of $\gamma$ black holes. A peculiarity of this solution is the appearance of a naked curvature singularity at $r=2m$ for some values of the parameter $\gamma$ (see for instance \cite{Kodama:2003ch}, \cite{Arrieta-Villamizar:2020brc}). The Zipoy-Vorhees metric, being a spheroidal deformation of a static spherically symmetric geometry can then represent a more realistic model for the exterior geometry of a static compact object. 

If one considers the presence of the electromagnetic fields in the vecinity or inside these compact objects then the situation is even more complicated, as expected. For the Weyl-Papapetrou ansatz, it has been long known that a transformation already exists that brings a static, axisymmetric vacuum solution to a non-trivial class of static solutions in Einstein-Maxwell theory \cite{weyl2}. In particular, the Schwarzschild solution can be transformed into  the Reissner-Nordstr\"{o}m  solution. This transformation is a special case of the Ehlers-Harrison transformations for the Ernst formalism \cite{ehlers-harrison} - \cite{Kinnersley} which map static vacuum solutions into static electrically charged Einstein-Maxwell solutions \cite{Weyl:1917gp}, \cite{Majumdar:1947eu}, \cite{Chng:2006gh} (see also \cite{Kleihaus:2009ff} in higher dimensions) and also transformations that map vacuum solutions into Einstein-Maxwell solutions with magnetic fields (see \cite{Ortaggio:2004kr,Yazadjiev:2005gs} and references there).

One should note at this point that some of the solution-generating techniques based on the Ernst formalism have been adapted to work for various other matter fields. Of particular interest is a technique introduced by Stephani in \cite{Stephani1} to generate perfect fluid solutions from vacuum solutions with various Killing symmetries. However, it turns out that not all equations of state are suitable for this sort of solution generating technique. Only two equations of state are compatible with this technique: $\rho=p$ for a spacelike Killing field and $\rho + 3p = 0$ for a timelike Killing field \cite{Garfinkle:1996ur}. In \cite{Krisch:2001ay} this technique was generalized to anisotropic fluids (see also \cite{Viaggiu:2006ft}). A similar technique was previously used in cosmological contexts in \cite{Belinski}. 

The main purpose of this paper is to show that some of these Ehlers-Harrison transformations can also be extended in presence of a fluid configuration. More specifically, starting from any interior solution (\ref{initialm}) sourced by the stress-energy tensor (\ref{initialf1}) one can easily generate by purely algebraic means the corresponding solutions in Einstein-Maxwell theory that correspond either to an electrically charged or to a magnetized interior solution. Just as the Ehlers-Harrison transformation will apply to every vacuum solution of the Einstein field equations, regardless of its physical significance, the transformations introduced in this paper will map any fluid solution (\ref{initialm}) sourced by a stress-energy tensor of the form (\ref{initialf1}) into a new exact solution of the Einstein-Maxwell-hydrodynamics solution. The generated solutions, being electrically charged or magnetized, can then be checked to satisfy the physical requirements for a viable physical model of a compact object if the original seed described by (\ref{initialm}) and (\ref{initialf1}) satisfies similar physical requirements.

The structure of this paper is as follows: in the next section we present the general equations of motion for the Einstein-Maxwell-hydrodynamics system. In section $3$ we introduce the map that generates the general electrically charged version of the metric (\ref{initialm}), solution of the above system. As an example of this procedure, we show how to straightforwardly obtain the electrically charged version of the Bowers and Liang solution \cite{Bowers} and present some of its properties. We also show how to obtain the electrically charged interior solution of the electrically charged Zipoy-Vorhees solution. In Section $4$ we construct the general magnetized version of the metric (\ref{initialm}) and construct the magnetized version of the interior Zipoy-Vorhees solution, which should correspond to an exterior magnetized Zipoy-Vorhees solution. In Section $5$ we comment on the junction conditions and the energy conditions for the solutions generated by our procedure. The final section contains a summary of our work and avenues for further work.

\section{The field equations in the Einstein-Maxwell-fluid system}

In GR the electromagnetic field is usually described by using the anti-symmetric Faraday tensor $F_{\mu\nu}=\nabla_{\mu}A_{\nu}-\nabla_{\nu}A_{\mu}$, where $A_{\mu}$ is the vector potential of the electromagnetic field. The Maxwell equations are then written as:
\beqs
\nabla_{\nu}(\star F)^{\mu\nu}&=&0,~~~~\nabla_{\nu}F^{\mu\nu}=4\pi J^{\mu},
\label{maxwell}
\eeqs
where $\star F_{\mu\nu}=\frac{1}{2}\epsilon_{\mu\nu\gamma\delta}F^{\gamma\delta}$ is the Hodge-dual tensor, while $\epsilon_{\mu\nu\gamma\delta}$ is the Levi-Civita tensor. Also, $J^{\mu}$ is the $4$-current that sources the electromagnetic field. The $4$-current $J^{\mu}$ can be further decomposed with respect to the fluid $4$-velocity $u^{\mu}$ as:
\beqs
J^{\mu}&=&\sigma_e u^{\mu}+j^{\mu},
\eeqs
where  $\sigma_e u^{\mu}$ is the convection current, while $j^{\mu}$ is the conduction current, such that $\sigma_e=-J^{\mu}u_{\mu}$ is the proper charge density. 

One can also define the electric $4$-vector $E_{\mu}$ and the magnetic $4$-vector $B_{\mu}$ by using the relations:
\beqs
E_{\mu}&=&F_{\mu\nu}u^{\nu}, ~~~~~B_{\mu}=-(\star F_{\mu\nu})u^{\nu}.
\label{elmagint}
\eeqs

The conduction current $j^{\mu}$ is related to the components of the electric field $E_{\mu}$ by means of Ohm's law. Our solution generating technique for the Einstein-Magneto-hydrodynamic (EMH) system will simply require a finite value for the conduction current $j^{\mu}$ and a vanishing electric field, therefore we will be working in the so-called ideal MHD approximation (see for instance \cite{Noh:2018skn}).

The electromagnetic stress-energy tensor, which enters the Einstein field equations is defined as:
\beqs
T_{\mu\nu}^{em}&=&\frac{1}{4\pi}\left(F_{\mu\gamma}F_{\nu}^{~\gamma}-\frac{1}{4}F_{\gamma\delta}F^{\gamma\delta}g_{\mu\nu}\right).
\eeqs

Then Einstein's field equations of the Einstein-Maxwell-fluid system can be written as:
\beqs
G_{\mu\nu}&=&8\pi T^{em}_{\mu\nu}+8\pi T_{\mu\nu}^{fluid},
\label{eqfinal}
\eeqs
where $G_{\mu\nu}$ is the Einstein tensor for the geometry (\ref{initialm}), while $T_{\mu\nu}^{fluid}$ is the stress-energy tensor (\ref{initialf1}). In absence of the fluid component these equations lead to the usual Einstein-Maxwell equations. The stress-energy conservation equations $T^{\mu\nu}_{~~;\nu}=0$ lead to the equations of motion for the fluid and they are a consequence of the Einstein equations, so we will not concern ourselves with them at this point.

For an electrically charged solution, in absence of a magnetic field, the only non-zero component of the electromagnetic potential is $A_t$, if the solution is static. On the other hand, the magnetic field of a magnetized star can have both toroidal and poloidal components. However, when one takes into account the toroidal components of the magnetic field then the spacetime geometry (\ref{initialm}) has to be modified and it must include other non-vanishing metric components \cite{gourgoulhon}. Therefore, in our work we shall assume that the toroidal components are null and the magnetic field is purely poloidal. In this case the only non-zero component of the electromagnetic potential is $A_{\varphi}$.

\section{The electrically charged model}

As is it well known by now, in absence of the fluid components, the Einstein-Maxwell equations can be simply solved starting with a static vacuum solution of the Einstein equations. This procedure is similar to some of the Ehlers-Harrison transformations in the Ernst formalism. For instance, starting with any static vacuum solution of Einstein's field equations one can generate the corresponding electrically charged solution as well as a solution involving a magnetic field (see for instance \cite{Chng:2006gh}-\cite{Yazadjiev:2005gs} and references there)\footnote{In the magnetic case the initial vacuum seed can be time-dependent, as long as $\frac{\partial}{\partial\varphi}$ is still a Killing vector in the seed geometry.}.

For example, let us start with the vacuum Zipoy-Vorhees solution given in (\ref{Zipoy}). Let us define the quantity $\Lambda=1-E_0^2 e^{-\psi}$, where $E_0$ is a constant parameter, which can be related to the charge parameter in the final solution. 

Then the charged version of the Zipoy-Vorhees metric (\ref{Zipoy}) takes the following form \cite{Chng:2006gh}, \cite{Richterek:2004bb} - \cite{Richterek:2002if} :
\beqs
ds^2&=&-\frac{e^{-\psi}}{\Lambda^2}dt^2+\Lambda^2e^{\psi}\big[e^{2\mu}(d\rho^2+dz^2)+\rho^2d\varphi^2\big],
\label{chargedZ}
\eeqs
while the electromagnetic potential takes the simple form:
\beqs
A_t&=&\frac{E_0e^{-\psi}}{\Lambda}.
\label{emz}
\eeqs
As it can be easily checked, this is an exact solution of the Einstein-Maxwell field equations (\ref{eqfinal}) and (\ref{maxwell}) with $J_{\mu}=0$. This is an expected result, since using the Ehlers-Harrison transformation any vacuum solution of the form (\ref{weyl}) will lead to a solution of the Einstein-Maxwell equations using (\ref{chargedZ}) and (\ref{emz}), as in \cite{weyl2}.

Our goal in this section is to generalize this Harrison transformation to more general axisymmetric solutions, which are sourced by fluids. However, one should expect that in absence of the fluid components the new solution generated by the solution generating technique presented here should simply reduce to the Einstein-Maxwell solution above. That is, for vacuum solutions, the generated Einstein-Maxwell fields take the form (\ref{chargedZ}) and (\ref{emz}): the geometry is modified by the electric field generated by (\ref{emz}) by means of the $\Lambda$ factors multiplying the metric components. However, simply by modifying the geometry as in (\ref{chargedZ}) to accommodate the presence of the electric field (\ref{emz}) one has to rescale accordingly the $4$-velocity of the fluid and the spacelike unit vectors in the radial and transverse directions by powers of $\Lambda$ to make them unit vectors in the new geometry (\ref{chargedZ}) with the electric field (\ref{emz}). This means that in the new geometry the original fluid quantities (such as the energy density and the radial and transverse pressures) should be modified by powers of $\Lambda$ as well. Also, there is another modification that one should expect for an electrically charged fluid: its energy density should receive a contribution $\rho_{el}$ due to electromagnetic energy of the fluid being charged in presence of an electric field. Finally, the Maxwell equations should also be modified since they are sourced now by a charged fluid as in (\ref{maxwell}). In conclusion, in presence of an electric field the original fluid quantities should be rescaled by powers of $\Lambda$ and the energy density $\rho$ should receive a contribution $\rho_{el}$, while the $4$-current contribution $J^{\mu}$ should be expressed using the proper charge density $\sigma_{e}$ in terms of the original uncharged fluid quantities.

 To see how this works in practice, let us consider the initial general geometry (\ref{initialm}). According to Einstein's field equations this geometry can be sourced in general by an anisotropic fluid having the stress-energy of the form (\ref{initialf1})\footnote{At this point we do not have to assume any physical requirements on the fluid source.}:
\beqs
T_{\mu\nu}^0&=&\rho^0u_{\mu}^0u_{\nu}^0+p_r^0\chi_{\mu}^0\chi_{\nu}^0+p_{\theta}^0\xi_{\mu}^0\xi_{\nu}^0+p_{\varphi}^0\zeta_{\mu}^0\zeta_{\nu}^0+2p_{r\theta}^0\chi_{(\mu}^0\xi_{\nu)}^0,
\label{initialf}
\eeqs
where $\rho^0$ is the fluid's energy density, $p_r^0$ is the radial pressure, while $p_{\theta}^0$, $p_{\varphi}^0$ and $p_{r\theta}^0$ are transverse components of the fluid pressure. Also $u_{\mu}^0=(-A, 0, 0, 0)$ is the $4$-velocity of the fluid, while $\chi_{\mu}^0=(0, B, 0, 0)$, $\xi_{\mu}^0=(0, 0, C, 0)$ and $\zeta_{\mu}^0=(0, 0, 0, D)$ are spacelike unit vectors in the radial and transverse directions.

Similar to the Harrison transformation one can construct now the following metric:
\beqs
ds^2&=&-\frac{A(r,\theta)^2}{\Lambda^2}dt^2+\Lambda^2\big[B(r,\theta)^2dr^2+C(r,\theta)^2d\theta^2+D(r,\theta)^2d\varphi^2\big],
\label{electricm}
\eeqs
where we defined $\Lambda=1-E_0^2A(r,\theta)^2$, with $E_0$ being a constant. This will be a solution of the Einstein-Maxwell-fluid equations (\ref{eqfinal}) together with the Maxwell equations (\ref{maxwell}) if the electromagnetic $4$-vector potential is $A_{\mu}=(A_t, 0, 0, 0)$ with
\beqs
A_t&=&\frac{E_0A(r,\theta)^2}{\Lambda},
\label{finalel}
\eeqs
while the fluid stress-energy tensor has the form:
\beqs
T_{\mu\nu}^{fluid}&=&\rho u_{\mu}u_{\nu}+p_r\chi_{\mu}\chi_{\nu}+p_{\theta}\xi_{\mu}\xi_{\nu}+p_{\varphi}\zeta_{\mu}\zeta_{\nu}+2p_{r\theta}\chi_{(\mu}\xi_{\nu)}.
\label{finalf}
\eeqs
Here we defined
\beqs
 \rho&=&\frac{\rho^0}{\Lambda^2}+\rho_{el},~~~p_r=\frac{p_r^0}{\Lambda^2},~~~p_{\theta}=\frac{p_{\theta}^0}{\Lambda^2},~~~p_{\varphi}=\frac{p_{\varphi}^0}{\Lambda^2},~~~p_{r\theta}=\frac{p_{r\theta}^0}{\Lambda^2}.
 \eeqs
Note that $u_{\mu}=\left(-\frac{A}{\Lambda}, 0, 0, 0\right)$ is the $4$-velocity of the charged fluid, while $\chi_{\mu}=(0, B\Lambda, 0, 0)$, $\xi_{\mu}=(0, 0, C\Lambda, 0)$ and $\zeta_{\mu}=(0, 0, 0, D\Lambda)$ are respectively spacelike unit vectors in the radial and the transverse angular directions computed in the final geometry (\ref{electricm}).
Finally, the contribution to the fluid energy density of the charge energy density is:
 \beqs
\rho_{el}&=&2(\rho^0+p_r^0+p_{\theta}^0+p_{\varphi}^0)\frac{E_0^2A(r,\theta)^2}{\Lambda^3}
\label{rhoel}
\eeqs 
 and the electric current is $J_{\mu}=(j_t, 0, 0, 0)$ where:
 \beqs
 j_t&=&-2(\rho^0+p_r^0+p_{\theta}^0+p_{\varphi}^0)\frac{E_0A(r,\theta)^2}{\Lambda^4},
\label{Jel}
 \eeqs
from which one can easily read the proper charge density $\sigma_e$.

It can be checked by brute force (for instance using Maple \cite{Maple}) that the fields given in (\ref{electricm}), (\ref{finalel}), (\ref{finalf}) provide an exact solution of the coupled Einstein-Maxwell-fluid system if (\ref{initialm}) and (\ref{initialf}) is an exact solution of the Einstein-fluid equations of motion. This will provide us with the generalization of the electric Harrison transformation in presence of an anisotropic fluid. 

Note that even if the electric potential can potentially be a function of the $\theta$ coordinate, if one computes the electric field using (\ref{elmagint}) one obtains purely electric fields and no magnetic fields, as expected. One should also note that the above transformations are direct generalizations of the transformations previously found and used in \cite{Stelea:2018shx} and \cite{Yazadjiev:2004bg} for metrics with spherical symmetry.

\subsection{The electrically charged version of the Bowers-Liang solution}

As an example of this solution-generating technique, let us consider first the charged version of the anisotropic Bowers-Liang solution. This solution, which was found by Bowers and Liang \cite{Bowers} corresponds to a anisotropic fluid with a homogeneous density distribution $\rho^0=constant$.  In their work they considered a spherically symmetric relativistic matter distribution and studied the behavior of such systems by incorporating the pressure anisotropy effects in the equation of the hydrostatic equilibrium. Their solution is given by (\ref{initialm}) where:
\beqs
A(r,\theta)^2&=&\bigg[\frac{3\left(1-\frac{2M}{R}\right)^{\frac{h}{2}}-\left(1-\frac{2m(r)}{r}\right)^{\frac{h}{2}}}{2}\bigg]^{\frac{2}{h}},~~~~B(r,\theta)^2=\frac{1}{1-\frac{2m(r)}{r}},~~~ C(r,\theta)=r, \\
D(r,\theta)&=&r\sin\theta,~~~\rho^0=\frac{3M}{4\pi R^3},~~~p_r^0=\rho^0\frac{\left(1-\frac{2m(r)}{r}\right)^{\frac{h}{2}}-\left(1-\frac{2M}{R}\right)^{\frac{h}{2}}}{3\left(1-\frac{2M}{R}\right)^{\frac{h}{2}}-\left(1-\frac{2m(r)}{r}\right)^{\frac{h}{2}}},\nonumber\\
\Delta^0&=&p_t^0-p_r^0=\frac{4\pi}{3}Cr^2\frac{(\rho^0+p_r^0)(\rho^0+3p_r^0)}{1-\frac{2m(r)}{r}},\nonumber
\eeqs
where $h=1-2C$, $m(r)=M\frac{r^3}{R^3}$ and $C$ is the anisotropy parameter. Note that for this solution $p_{\theta}^0=p_{\varphi}^0\equiv p_t^0$, while $p_{r\theta}^0=0$ in (\ref{initialf}).

Then, using the results from the previous section, the electrically charged Bowers-Liang solution will simply be given by (\ref{electricm}) supplemented by (\ref{finalel}) and (\ref{finalf}). The final geometry is still spherically symmetric. For $C=0$ one obtains the electrically charged interior Schwarzschild solution discussed in \cite{Yazadjiev:2004bg} in a slightly different form, in absence of the dilaton field.  

Note that the anisotropy factor becomes in our case:
\beqs
\Delta&=&p_t-p_r=\frac{\Delta_0}{\Lambda^2}=\frac{4\pi}{3}\frac{Cr^2}{\Lambda^2}\frac{(\rho^0+p_r^0)(\rho^0+3p_r^0)}{1-\frac{2m(r)}{r}}
\eeqs

Since in origin $r=0$ then $\Delta=0$ as expected. Also, since
\beqs
\Lambda_0=1-E_0^2\left(\frac{3\left(1-\frac{2M}{R}\right)^{\frac{h}{2}}-1}{2}\right)^{\frac{2}{h}},
\eeqs 
then in the electrically charged Bowers-Liang solution the radial pressure in origin becomes:
\beqs
p_r(0)&=&\frac{\rho^0}{\Lambda_0^2}\frac{1-\left(1-\frac{2M}{R}\right)^{\frac{h}{2}}}{3\left(1-\frac{2M}{R}\right)^{\frac{h}{2}}-1},
\eeqs
and the critical value of the quantity $\frac{2M}{R}$ for which the central pressure becomes infinite is\footnote{For this value $\Lambda_0\ra 1$, there is no physical critical value of $\frac{2M}{R}$ for which $\Lambda_0=0$.}:
\beqs
\frac{2M}{R}|_{cr}&=&1-\left(\frac{1}{3}\right)^{\frac{2}{h}}.
\label{crmass}
\eeqs
The critical value of the ratio $\frac{2M}{R}$ is the same as the critical value of the original Bowers and Liang uncharged solution.

 In this particular case, as it is well-known, the original Bowers-Liang solution can be smoothly matched to the exterior Schwarzschild vacuum solution at the surface $r=R$ where the radial pressure vanishes $p_r(R)=0$. Correspondingly, the charged version of the Bowers-Liang solution should be smoothly matched to the corresponding electrically charged exterior solution, which should correspond to the Reissner-Nordstr\"om solution. Indeed, in our coordinates the charged exterior geometry corresponds to:
\beqs
ds^2&=&-\frac{1-\frac{2M}{r}}{\Lambda^2}+\frac{\Lambda^2}{1-\frac{2M}{r}}dr^2+\Lambda^2r^2(d\theta^2+\sin^2\theta d\varphi^2),
\label{elSchw}
\eeqs
where $\Lambda=1-E_0^2\left(1-\frac{2M}{r}\right)$, while the electromagnetic potential has the only non-zero component:
 \beqs
A_t=\frac{E_0\left(1-\frac{2M}{r}\right)}{\Lambda}.
\eeqs
This is the solution obtained by applying our solution-generating technique directly on the vaccum exterior Schwarzschild geometry. By performing the coordinate transformation $\tilde{r}=\Lambda r$ and rescaling  the time coordinate such that $\tilde{t}=\frac{t}{1-E_0^2}$ then this exterior geometry reduces to the usual Reissner-Nordstr\"om geometry with mass $M_{RN}=M(1+E_0^2)$ and charge $Q_{RN}=2ME_0$, as expected.

It should be clear by now that all the junction conditions are smoothly satisfied at $r=R$ and that the charged interior Bowers-Liang solution is smoothly connected to the exterior geometry. To see how this works in practice, let us recast the electrically charged Bowers-Liang solution in the more familiar form:
\beqs
ds^2&=&-\frac{f(r)}{\Lambda(r)^2}dt^2+\Lambda(r)^2\bigg[\frac{dr^2}{g(r)}+r^2(d\theta^2+\sin^2\theta d\varphi^2)\bigg],
\label{EBL}
\eeqs
where $\Lambda(r)=1-E_0^2f(r)$, with
\beqs
f(r)&=&\bigg[\frac{3\left(1-\frac{2M}{R}\right)^{\frac{h}{2}}-\left(1-\frac{2m(r)}{r}\right)^{\frac{h}{2}}}{2}\bigg]^{\frac{2}{h}},~~~g(r)=1-\frac{2Mr^2}{R^3}.
\eeqs
Note that when $r=R$ one has $f(R)=g(R)=1-\frac{2M}{R}$.
In this case, if the exterior geometry is the Schwarzschild solution then the smooth matching of the charged interior Bowers-Liang to the exterior charged geometry require that the first and the second fundamental forms are continuous on the junction surface $r=R$. If $\chi^a=(0,\frac{\sqrt{g(r)}}{\Lambda(r)},0,0)$ is the unit vector in the radial direction in the electrically charged metric (\ref{EBL}) then the induced metric on the surface $r=R$ is found to be:
\beqs
h_{ab}dx^adx^b&=&-\frac{1-\frac{2M}{R}}{\Lambda^2}+\Lambda^2R^2(d\theta^2+\sin^2\theta d\varphi^2),
\eeqs
which is the same with the one induced from the charged exterior geometry in (\ref{elSchw}) on the same surface $r=R$. Now, the second fundamental form corresponds to the extrinsec curvature $K_{cd}=h^a_ch^b_d\chi_{a;b}$ and it has the nonzero components:
\beqs
K^t_t&=&\frac{\sqrt{g(r)}}{2\Lambda(r)^2f(r)}\left(\Lambda(r)f'(r)-2f(r)\Lambda'(r)\right),~~~K^{\theta}_{\theta}=K^{\varphi}_{\varphi}=\frac{\sqrt{g(r)}}{r\Lambda(r)^2}\left(r\Lambda'(r)-\Lambda\right),
\eeqs
where ' denotes here the differentiation with respect to $r$ and $\Lambda=1-E_0^2f(r)$. Since on the surface $r=R$ one has $f(R)=1-\frac{2M}{R}$ and its derivative $f'(r=R)=\frac{2M}{R^2}$, it is now clear that the second fundamental form is also continuous on the hypersurface $r=R$ and therefore the electrically charged Bowers-Liang solution will smoothly match the electrically charged exterior solution (\ref{elSchw}) on the surface $r=R$.

Finally, if one takes $h=0$ in the electrically charged Bowers-Liang solution one obtains the charged version of the so-called Florides solution \cite{florides}. It corresponds to an anisotropic object with zero radial pressure $p_r=0$, which is sustained only by tangential stresses. 

\subsection{The charged Zipoy-Vorhees interior solution}

It is well known that the exterior geometry of a compact spinning object is not generically described by the Kerr geometry since one expects that the multipole moments structure of a compact object can be more general than that of a Kerr black hole. Recetly, in \cite{Raposo:2018xkf} it was developed a general-relativistic framework to construct vacuum geometries that are perturbative deviations from the spherically symmetric Schwarzschild geometry. That class of perturbative solutions was obtained by solving the vacuum Einstein field equations order by order in a small multipole moment expansion and this perturbative solution is expected to be useful to parameterize the exterior axisymmetric geometry of a compact object. While the analysis of \cite{Raposo:2018xkf} was concerned only with the exterior vacuum geometry, this approach was further extended in \cite{Raposo:2020yjy} to include matter fields, in particular, the authors in \cite{Raposo:2020yjy} considered the effects of a perfect fluid in the interior of the compact object. To this end, the interior geometry was found numerically and the matching with the exterior geometry from \cite{Raposo:2018xkf} was done by taking into account the junction conditions at the boundary with the external solution.

In this work we focus on a somewhat different approach, as we will consider static axisymmetric fluid geometries from the beginning. As such, if one considers the quadrupole deviation from the Schwarzschild geometry, then taking the Zipoy-Vorhees geometry as the static exterior solution is the natural step. The search for the full interior solution of the Zipoy-Vorhees has a long history and it is still an active area of research \cite{Abishev:2019okd} - \cite{Herrera:2003wh}. For our purposes we shall make use of one of the solutions found in \cite{Stewart1982}. More specifically, we shall use the second interior solution in \cite{Stewart1982}, which was based on the modified Adler interior solution. The interior geometry takes then the form:
\beqs
ds^2&=&-f(r)^{2\gamma}dt^2+f(r)^{2(1-\gamma)}\Delta(r)^{\gamma^2-2}\Sigma(r,\theta)^{1-\gamma^2}dr^2+r^2f(r)^{2\gamma(\gamma-1)}\Phi(r,\theta)^{1-\gamma^2}d\theta^2 \nonumber \\&&+r^2f(r)^{2(1-\gamma)}\sin^2\theta d\varphi^2,
\label{AZV}
\eeqs
where one defined:
\beqs
f(r)&=&A+Br^2,~~~\Sigma(r,\theta)=1+\frac{Cr^2}{(A+3Br^2)^{\frac{2}{3}}}+\frac{1}{4}\frac{C^2r^4}{(A+3Br^2)^{\frac{4}{3}}}\sin^2\theta, \nonumber\\
\Delta(r)&=&1+\frac{Cr^2}{(A+3Br^2)^{\frac{2}{3}}},~~~\Phi(r,\theta)=(A+Br^2)^2+\frac{C^2r^4V(r)}{4(A+3Br^2)^{\frac{4}{3}}}\sin^2\theta,
\eeqs 
while
\beqs
V(r)&=&1+\frac{6}{a}\frac{1-\frac{5m}{2a}}{1-\frac{m}{a}}(a-r),~~~A=\frac{1-\frac{5m}{2a}}{\left(1-\frac{2m}{a}\right)^{\frac{1}{2}}},~~~B=\frac{m}{2a^3\left(1-\frac{2m}{a}\right)^{\frac{1}{2}}},~~~C=\frac{2m\left(1-\frac{m}{a}\right)^{\frac{2}{3}}}{a^3\left(1-\frac{2m}{a}\right)^{\frac{1}{3}}}.\nonumber
\eeqs
The interior solution matches the exterior Zipoy-Vorhees geometry on the boundary surface $r=a$ and all the regularity conditions are satisfied there \cite{Stewart1982}. One can compute the expressions of the fluid quantities (the energy density $\rho^0$ and the corresponding pressures $p_r^0$, $p_{\theta}^0$, $p_{\varphi}^0$ and $p_{r\theta}^0$) however, they are lengthy and not particularly illuminating. To extract some physical intuition one has to make a few approximations: if one defines $\gamma=1+\epsilon$ and $\delta=\frac{m}{a}$, then, up to the order $\epsilon\delta, \delta^2$ one obtains\footnote{Note that there are some typos in the corresponding expressions in \cite{Stewart1982}.}:
\beqs
\rho^0&=&\frac{1}{8\pi}\bigg[\frac{6\delta(1+\epsilon)}{a^2}-\frac{20r^2\delta^2}{a^4}\bigg],~~~p_r^0=\frac{1}{8\pi}\bigg[\frac{6\delta^2}{a^2}-\frac{2r^2\delta^2}{a^4}\bigg],\nonumber\\
p_{\theta}^0&=&p_{\varphi}^0=\frac{1}{8\pi}\bigg[\frac{6\delta^2}{a^2}-\frac{r^2\delta^2}{a^4}\bigg], 
\label{paramrho}
\eeqs
while to this order one has $p_{r\theta}^0=0$. Note that one has $p_r^0(r=0)=p_{\theta}^0(r=0)=p_{\varphi}^0(r=0)=\frac{6\delta^2}{a^2}$, while the energy density $\rho^0$ and the radial pressure $p_r^0$ decrease monotonically with $r$:
\beqs
\frac{d\rho^0}{dr}=-\frac{40\delta^2r}{8\pi a^4}<0, ~~~\frac{dp_r^0}{dr}=-\frac{4\delta^2r}{8\pi a^4}<0.
\eeqs

One is now ready to present the electrically charged interior Zipoy-Vorhees solution. Let us define the quantity $\Lambda=1-E_0^2f(r)^{2\gamma}$, where $E_0$ is a constant that will be related lately to the total charge of the solution. Then the metric of the interior charged fluid becomes:
\beqs
ds^2&=&-\frac{f(r)^{2\gamma}}{\Lambda^2}dt^2+\Lambda^2\bigg[f(r)^{2(1-\gamma)}\Delta(r)^{\gamma^2-2}\Sigma(r,\theta)^{1-\gamma^2}dr^2+r^2f(r)^{2\gamma(\gamma-1)}\Phi(r,\theta)^{1-\gamma^2}d\theta^2 \nonumber \\&&+r^2f(r)^{2(1-\gamma)}\sin^2\theta d\varphi^2\bigg],
\label{EAZV}
\eeqs
while the electric potential is given by $A_t=\frac{E_0f(r)^{2\gamma}}{\Lambda}$. This geometry and electric potential will provide us with an electrically charged solution of the Einstein-Maxwell-fluid equations (\ref{eqfinal}) and (\ref{maxwell}) if the fluid stress-energy tensor has the form given in (\ref{finalf}), while the electric current takes the form (\ref{Jel}).

Note that as shown in Section $5$ this interior solution will smoothly match the exterior charged Zipoy-Vorhees solution given in (\ref{chargedZ}). After a rescaling of the time coordinate one can compute the total mass $M=\gamma m (1+E_0^2)$ and the total charge $Q=2\gamma m E_0$ of the charged Zipoy-Vorhees geometry.

Furthermore, one can check that up to the order $\epsilon\delta, \delta^2$ one obtains:
\beqs
\rho&=&\frac{27}{\pi a^4(1-E_0^2)^4}\bigg[\frac{(1+\epsilon)a^2\delta}{36}-\frac{5\delta^2r^2}{54}-\left(\frac{a^2\delta}{6}(3\delta+2(1+\epsilon))-\frac{11r^2\delta^2}{9}\right)\frac{E_0^2}{3}\nonumber\\
&&+\left(\frac{2\delta^2}{3}+\frac{\delta a^2(1+\epsilon)}{12}-\frac{13r^2\delta^2}{27}\right)E_0^4\bigg],~~~
p_r=\frac{1}{8\pi}\bigg[\frac{6\delta^2}{(1-E_0^2)^2a^2}-\frac{2r^2\delta^2}{(1-E_0^2)^2a^4}\bigg],\nonumber\\ p_{\theta}&=&p_{\varphi}=\frac{1}{8\pi}\bigg[\frac{6\delta^2}{(1-E_0^2)^2a^2}-\frac{r^2\delta^2}{(1-E_0^2)^2a^4}\bigg],~~~p_{r\theta}=0.
\eeqs
For $E_0=0$ these expressions reduce to those corresponding in the uncharged case (\ref{paramrho}). Note that the value $E_0=1$ corresponds to the extremaly charged solution for which $M=Q$ and one should restrict the values of the parameter $E_0$ to never reach this value.

One can check that up to this order of approximation the density is once again decreasing with the radius, as well as the radial pressure:
\beqs
\frac{d\rho}{dr}&=&-\frac{r\delta^2(26E_0^4-22E_0^2+5)}{\pi a^4(1-E_0^2)^4}<0, ~~~\frac{dp_r}{dr}=-\frac{\delta^2r}{\pi a^4(1-E_0)^2}<0.
\eeqs
Finally, all the pressures have the same value in origin $r=0$ as expected:
\beqs
p_r(0)&=&p_{\theta}(0)=p_{\varphi}(0)=\frac{3\delta^2}{4\pi a^2(1-E_0^2)^2}.
\eeqs
This completes the derivation of the charged Zipoy-Vorhees interior solution.

\section{The magnetized solution}

As it was previously shown in \cite{Richterek:2004bb} - \cite{Richterek:2002if}, starting with the original Zipoy-Vorhees solution it is also possible to obtain its magnetized version by using a magnetizing Harrison transformation:
\beqs
ds^2&=&\Lambda^2\bigg[-e^{-\psi}dt^2+e^{\psi}\big[e^{2\mu}(d\rho^2+dz^2)\big]\bigg]+\frac{\rho^2e^{-\psi}}{\Lambda^2}d\varphi^2,
\label{MZV}
\eeqs
where now we denote $\Lambda=1+B_0^2\rho^2e^{-\psi}$ and the electromagnetic potential has the only nonzero component $A_{\varphi}=\frac{B_0\rho^2e^{-\psi}}{\Lambda}$. Here $B_0$ is a constant. This geometry and electromagnetic potential provide us with a solution of the Einstein-Maxwell equations. As it turns out, one can generalize this magnetizing Harrison transformation in presence of more general fluid interior solutions with axial symmetry.

Similarly to the electrically charged case, given a general solution (\ref{initialm}) sourced by the stress-energy tensor (\ref{initialf}) of the Einstein-fluid field equations, one can write down directly the corresponding magnetized solution in the following form:
\beqs
\label{finalmagm}
ds^2&=&\Lambda^2\big[-A(r,\theta)^2dt^2+B(r,\theta)^2dr^2+C(r,\theta)^2d\theta^2\big]+\frac{D(r,\theta)^2}{\Lambda^2}d\varphi^2,\\
A_{\varphi}&=&\frac{B_0D(r,\theta)^2}{\Lambda},~~~\Lambda=1+B_0^2D(r,\theta)^2,\nonumber
\eeqs
where the stress-energy tensor of the anisotropic fluid is given by:
\beqs
T_{\mu\nu}^{fluid}&=&\rho u_{\mu}u_{\nu}+p_r\chi_{\mu}\chi_{\nu}+p_{\theta}\xi_{\mu}\xi_{\nu}+p_{\varphi}\zeta_{\mu}\zeta_{\nu}+2p_{r\theta}\chi_{(\mu}\xi_{\nu)},
\label{finalfmag}
\eeqs
with
\beqs
 \rho&=&\frac{\rho^0}{\Lambda^2},~~~p_r=\frac{p_r^0}{\Lambda^2},~~~p_{\theta}=\frac{p_{\theta}^0}{\Lambda^2},~~~p_{\varphi}=\frac{p_{\varphi}^0}{\Lambda^2}+\sigma_m,~~~p_{r\theta}=\frac{p_{r\theta}^0}{\Lambda^2},
 \label{magfluid}
 \eeqs
 while:
 \beqs
 \sigma_m&=&-2(\rho^0-p_r^0-p_{\theta}^0+p_{\varphi}^0)\frac{B_0^2D(r,\theta)^2}{\Lambda^3}
\label{sigm}
 \eeqs
and the only non-vanishing component of the $4$-current $J_{\mu}$ is:
\beqs
 J_{\varphi}&=&2(\rho^0-p_r^0-p_{\theta}^0+p_{\varphi}^0)\frac{B_0D(r,\theta)^2}{\Lambda^4}.
\label{Jmag}
 \eeqs

Note that in this case the electric field is null $E_{\mu}=0$ and our results apply within  the ideal MHD approximation. Furthermore, the components of the magnetic field can be easily computed in the form:
\beqs
B_r&=&-\frac{2B_0\partial_{\theta}D(r,\theta)}{\Lambda r}, ~~~ B_{\varphi}=\frac{2B_0r\partial_rD(r,\theta)}{\Lambda},
\label{polB}
\eeqs
confirming that the magnetic field is poloidal in nature.

Finally, in the above geometry we used $u_{\mu}=\left(-A\Lambda, 0, 0, 0\right)$ as the $4$-velocity of the fluid, while $\chi_{\mu}=(0, B\Lambda, 0, 0)$, $\xi_{\mu}=(0, 0, C\Lambda, 0)$ and $\zeta_{\mu}=(0, 0, 0, \frac{D}{\Lambda})$ are respectively spacelike unit vectors in the radial and the transverse angular directions.

This solution is a direct generalization of the corresponding transformations considered previously in \cite{Yazadjiev:2011ks} and \cite{Stelea:2018cgm} for solutions with spherical symmetry to a more general class of interior solutions with axial symmetry. 

\subsection{A magnetized interior axially-symmetric solution}

Using a particular case of the above magnetizing technique, we already discussed the magnetized version of the Bowers-Liang solution in a previous work \cite{Stelea:2018cgm}. There it was found that in the magnetized version of the Bowers-Liang solution the central pressure becomes infinite for the same critical mass (\ref{crmass}) as in the original Bowers-Liang solution. Moreover, the magnetized interior solution matches smoothly the exterior Schwarzschild-Melvin solution.

As another simple example, we shall consider here the solution found in section $V$ in \cite{Herrera:2013hm}. This solution describes an incompressible isotropic spheroid, whose metric is given by:
\beqs
ds^2&=&-\left(\frac{\alpha r^2+\beta+a r \cos\theta}{\gamma r^2+\delta+b r \cos\theta}\right)^2dt^2+\frac{1}{\left(\gamma r^2+\delta+b r \cos\theta\right)^2}\bigg[dr^2+r^2\left(d\theta^2+\sin^2\theta d\varphi^2\right)\bigg],\nonumber
\eeqs
which is sourced by a fluid with isotropic pressures and a homogenous energy density:
\beqs
\rho^0&=&\frac{12\gamma\delta-3b^2}{8\pi},\nonumber\\
p_r^0&=&p_{\theta}^0=p_{\varphi}^0=\frac{3b^2-12\gamma\delta}{8\pi}\bigg[1-\frac{2(2\alpha\delta+2\beta\gamma-ab)}{12\gamma\delta-3b^2}\frac{\gamma r^2+\delta+b r \cos\theta}{\alpha r^2+\beta+a r \cos\theta}\bigg],\nonumber\\
p_{r\theta}^0&=&0.
\eeqs
If one defines:
\beqs
\Lambda&=&1+\frac{B_0^2r^2\sin^2\theta}{\left(\gamma r^2+\delta+b r \cos\theta\right)^2},
\eeqs
then the magnetized version of the incompressible isotropic spheroid becomes:
\beqs
ds^2&=&\Lambda^2\bigg[-\left(\frac{\alpha r^2+\beta+a r \cos\theta}{\gamma r^2+\delta+b r \cos\theta}\right)^2dt^2+\frac{1}{\left(\gamma r^2+\delta+b r \cos\theta\right)^2}\left(dr^2+r^2d\theta^2\right)\bigg]\nonumber\\
&&+\frac{r^2\sin^2\theta}{\Lambda^2\left(\gamma r^2+\delta+b r \cos\theta\right)^2}d\varphi^2,
\eeqs
while the magnetic potential is:
\beqs
A_{\varphi}&=&\frac{B_0r^2\sin^2\theta}{\Lambda}
\eeqs
and the magnetized components of the fluid energy density and pressures can be computed easily from (\ref{magfluid}). One should note that the effective transverse pressure along the transverse $\varphi$ direction becomes $p_{\varphi}=\frac{p_{\varphi}^0}{\Lambda^2}+\sigma_m$, where $\sigma_m$ is in our case:
\beqs
\sigma_m&=&-2(\rho^0-p_r^0-p_{\theta}^0+p_{\varphi}^0)\frac{B_0^2r^2\sin^2\theta}{\Lambda^3}.
\eeqs
We should mention that we have checked in Maple that this is indeed an exact solution of the Einstein-Maxwell-fluid equations. As expected, there is now manifest an anisotropy in the pressure distribution, due to the presence of the magnetic field. Unfortunately, there is no known exterior solution in the Weyl class that could be matched to this interior solution \cite{Herrera:2013hm}, so its physical relevance is lacking at this point.

\subsection{The magnetized interior Zipoy-Vorhees solution}

Let us consider now the magnetized version of the Zipoy-Vorhees interior solution (\ref{AZV}). Starting with the modified Adler interior solution (\ref{AZV}) one can write down directly the magnetized geometry as:
\beqs
ds^2&=&\Lambda^2\bigg[-f(r)^{2\gamma}dt^2+f(r)^{2(1-\gamma)}\Delta(r)^{\gamma^2-2}\Sigma(r,\theta)^{1-\gamma^2}dr^2+r^2f(r)^{2\gamma(\gamma-1)}\Phi(r,\theta)^{1-\gamma^2}d\theta^2\bigg] \nonumber \\&&+\frac{r^2f(r)^{2(1-\gamma)}\sin^2\theta}{\Lambda^2} d\varphi^2,
\label{MAZV}
\eeqs
where we defined $\Lambda=1+B_0^2r^2f(r)^{2(1-\gamma)}\sin^2\theta$. The only nonzero component of the electromagnetic potential is:
\beqs
A_{\varphi}&=&\frac{B_0r^2f(r)^{2(1-\gamma)}\sin^2\theta}{\Lambda}.
\eeqs
If the original interior Zipoy-Vorhees solution is sourced by an anisotropic fluid with the stress-energy of the form (\ref{initialf}), then the magnetized version of the fluid components has the stress-energy described by (\ref{finalfmag}), with components given by (\ref{magfluid}) and (\ref{sigm}). Finally, the $4$-current which sources the Maxwell equations has the only nonzero component (\ref{Jmag}).

This interior solution will smoothly match the exterior solution, which in our case is represented by the magnetized Zipoy-Vorhees given in (\ref{MZV}), as discussed in Section $5$. The components of the poloidal magnetic field can be easily computed using (\ref{polB}).

Let us define again $\gamma=1+\epsilon$ and $\delta=\frac{m}{a}$. Then, up to the orders $\epsilon\delta$, $\delta$ one obtains:
\beqs
\rho&=&\frac{1}{8\pi(1+B_0^2r^2\sin^2\theta)^2}\bigg[\frac{6\delta(1+\epsilon)}{a^2}-\frac{20\delta^2r^2}{a^4}\bigg],~~~p_r=\frac{1}{8\pi(1+B_0^2r^2\sin^2\theta)^2}\bigg[\frac{6\delta^2}{a^2}-\frac{2\delta^2r^2}{a^4}\bigg],\nonumber\\
p_{\theta}&=&\frac{1}{8\pi(1+B_0^2r^2\sin^2\theta)^2}\bigg[\frac{6\delta^2}{a^2}-\frac{\delta^2r^2}{a^4}\bigg],\nonumber\\
p_{\varphi}&=&\frac{1}{8\pi(1+B_0^2r^2\sin^2\theta)^2}\bigg[\frac{6\delta^2}{a^4}-\frac{\delta^2r^2}{a^4}+2B_0^2r^2\sin^2\theta\left(\frac{17\delta^2r^2}{a^4}+\frac{6\delta(2\delta-(1+\epsilon))}{a^2}\right)\nonumber\\&&+B_0^4r^4\sin^4\theta\left(\frac{35\delta^2r^2}{a^4}+\frac{6\delta(3\delta-2(1+\epsilon))}{a^4}\right)\bigg],~~~p_{r\theta}=0.
\eeqs
 The anisotropy induced by the magnetic field is now manifest, as $p_{\varphi}$ is clearly modified away by $\sigma_m$ and it is different now from the other tangential pressure $p_{\theta}$. Note that for $B_0=0$ one re-obtains the values corresponding to the original interior Zipoy-Vorhees (\ref{paramrho}).

One can also show that up to this order the density and the radial pressure decrease monotonically with $r$:
\beqs
\frac{d\rho}{dr}&=&-\frac{4B_0^2r\sin^2\theta}{(1+B_0^2r^2\sin^2\theta)^3}\bigg[\frac{6\delta(1+\epsilon)}{a^2}-\frac{20\delta^2r^2}{a^4}\bigg]-\frac{40\delta^2r}{a^4(1+B_0^2r^2\sin^2\theta)^2}<0,\nonumber\\
\frac{dp_r}{dr}&=&-\frac{8\delta^2}{a^4(1+B_0^2r^2\sin^2\theta)^2}\left(2+B_0^2r^2\sin^2\theta+3a^2B_0^2\sin^2\theta\right)<0,
\eeqs
as expected.

\section{A note on the junction and energy conditions}

Consider a static interior solution with axial symmetry in GR. Let us suppose that its metric is given by (\ref{initialm}) and that it can be smoothly matched to a vacuum exterior solution in the Weyl-Papapetrou class on a surface $r=R$, so that it represents a valid interior solution. The boundary junction conditions that require that the first and the second fundamental forms be continuous at the boundary are equivalent to requiring that the functions $A(r,\theta)^2$, $C(r,\theta)^2$ and $D(r,\theta)^2$ are all continuous and have continuous first derivatives at $r=R$. Moreover the functions $B(r,\theta)^2$ and its derivative $(B(r,\theta)^2)_{,\theta}$ must be continuous as well.

Indeed, for a geometry of the form (\ref{initialm}), since the matching to an exterior geometry is done on the hypersurface $r=R$ then one has to compute the induced metric on this surface as well as the components of the extrinsic curvature. Defining the unit vector in the radial direction by $\chi^a=(0,\frac{1}{B(r,\theta)},0,0)$ then the induced metric on a surface of constant radius is $h_{ab}=g_{ab}-\chi_a\chi_b$, that is:
\beqs
h_{ab}dx^adx^b&=&-A(r,\theta)^2dt^2+C(r,\theta)^2d\theta^2+D(r,\theta)^2d\varphi^2.
\eeqs
If the first fundamental form is to be continuous with the one induced from the exterior geometry on a surface of constant radius $r=R$ then the metric components $A(r,\theta)$, $C(r,\theta)$ and $D(r,\theta)$ have to be continuous at $r=R$ with the corresponding values from the exterior geometry.

To see how this is working in practice, let us consider a general exterior vacuum geometry (while we are still matching it to the interior geometry on a surface of constant radius $r=R$). Let us describe the vacuum exterior geometry using the general static metric of the form:
\beqs
ds^2_{ext}&=&-\tilde{A}(r,\theta)^2dt^2+\tilde{B}(r,\theta)^2dr^2+\tilde{C}(r,\theta)^2d\theta^2+\tilde{D}(r,\theta)^2d\varphi^2.
\label{initialmv}
\eeqs
The exterior geometry induced on the same hypersurface $r=R$ will then be:
\beqs
\tilde{h}_{ab}dx^adx^b&=&-\tilde{A}(r,\theta)^2dt^2+\tilde{C}(r,\theta)^2d\theta^2+\tilde{D}(r,\theta)^2d\varphi^2.
\eeqs
Therefore, the continuity of the metric components induced by the two geometries on the hypersurface $r=R$ implies $A(R,\theta)=\tilde{A}(R,\theta)$, $C(R,\theta)=\tilde{C}(R,\theta)$ and $D(R,\theta)=\tilde{D}(R,\theta)$, as expected.

Turning now our attention to the components of the second fundamental form, \textit{aka} the extrinsic curvature tensor $K_{ab}=h^c_a h^d_b\chi_{d;c}$ then one obtains for the interior metric (\ref{initialm}):
\beqs
K^t_t&=&\frac{A_{,r}}{AB},~~~K^{\theta}_{\theta}=\frac{C_{,r}}{CB},~~~K^{\varphi}_{\varphi}=\frac{D_{,r}}{DB}.
\eeqs
Correspondingly, for the extrinsic components of the exterior vacuum geometry one has:
\beqs
\tilde{K}^t_t&=&\frac{\tilde{A}_{,r}}{\tilde{A}\tilde{B}},~~~\tilde{K}^{\theta}_{\theta}=\frac{\tilde{C}_{,r}}{\tilde{C}\tilde{B}},~~~\tilde{K}^{\varphi}_{\varphi}=\frac{\tilde{D}_{,r}}{\tilde{D}\tilde{B}},
\eeqs
where $\tilde{K}_{ab}=\tilde{h}^c_a\tilde{h}^d_b\tilde{\chi}_{d;c}$ is the extrinsic curvature of the hypersurface $r=R$ in the exterior geometry, with $\tilde{\chi}^a=(0, \frac{1}{\tilde{B}(r,\theta)},0,0)$ being the unit radial vector in the exterior vacuum geometry.

If the second fundamental form is to be continuous on the matching surface $r=R$ then it should be obvious that one must require the functions $B(r,\theta)$, $C(r,\theta)$ and $D(r,\theta)$ to be continuous at $r=R$, as well as the values of the partial derivatives $A_{,r}$, $C_{,r}$ and $D_{,r}$ to be equal to the corresponding tilded values obtained from the exterior geometry. This means that we have to require $A(R,\theta)=\tilde{A}(R,\theta)$, $B(R,\theta)=\tilde{B}(R,\theta)$, $C(R,\theta)=\tilde{C}(R,\theta)$ and $D(R,\theta)=\tilde{D}(R,\theta)$, as well as $A_{,r}(R,\theta)=\tilde{A}_{,r}(R,\theta)$,  $C_{,r}(R,\theta)=\tilde{C}_{,r}(R,\theta)$ and  $D_{,r}(R,\theta)=\tilde{D}_{,r}(R,\theta)$.

However, if the matching to the exterior geometry is not smooth on the junction surface $r=R$ then this will mean that some components of the extrinsic curvature will not equal the corresponding components of the extrinsic curvature of the exterior geometry when evaluated on the same surface. In this case, using the Israel-Lanczos junction formalism \cite{Israel:1966rt}, \cite{lanczos} if one defines the quantity $k_{ab}=K_{ab}-\tilde{K}_{ab}$, then the Lanczos stress-energy tensor on the surface $r=R$ is defined as:
\beqs
S^a_b&=&-\frac{1}{8\pi}\left(k^a_b-\delta^a_b k^c_c\right).
\label{Lanczos}
\eeqs
Discontinuities in the components of the extrinsic curvature on the junction surface $r=R$ will lead to nonzero components of the tensor $k_{ab}$ and correspondingly non-zero components of the superficial stress-energy tensor (\ref{Lanczos}) on that surface.

As it was originally shown in \cite{Stewart1982}, these conditions for smooth matching are satisfied for the interior Zipoy-Vorhees solutions based on the modified Adler solution, while the exterior vacuum solution is in this case the vacuum Zipoy-Vorhees solution (\ref{Zipoy}). Indeed, in this case the matching is done on the surface $r=a$. One can easily check that on this surface one has:
\beqs
A(a,\theta)&=&-(a-2m)^{\gamma}a^{-\gamma},\nonumber\\
B(a,\theta)&=&a^{\gamma^2+\gamma-1}(a-2m)^{\gamma^2-\gamma-1}(m^2\sin^2\theta+a^2-2ma)^{1-\gamma^2},\nonumber\\
C(a,\theta)&=&a^{\gamma^2+\gamma}(a-2m)^{\gamma^2-\gamma}(m^2\sin^2\theta+a^2-2ma)^{1-\gamma^2},\nonumber\\
D(a,\theta)&=&a^{1+\gamma}(a-2m)^{1-\gamma}\sin^2\theta,
\label{firstform}
\eeqs
using the interior modified Adler solution from (\ref{AZV}) and they do agree with the metric components induced by the exterior Zipoy-Vorhees metric on a surface of constant radius $r=a$.

To check that the interior solution (\ref{AZV}) will smoothly match the exterior Zipoy-Vorhees solution one has to check the continuity of the components of the extrinsic curvature on the junction surface $r=a$. As we have previously seen, this amounts to check the continuity of the radial derivatives of the metric components $A(r,\theta)$, $C(r,\theta)$ and $D(r,\theta)$ on $r=a$. After some tedious computations, one explicitly obtains for the interior solution:
\beqs
A_{,r}(a,\theta)&=&-2m\gamma a^{-1-\gamma}(a-2m)^{\gamma-1},\nonumber\\
C_{,r}(a,\theta)&=&2a^{\gamma^2+\gamma-1}(m^2\sin^2\theta+a^2-2ma)^{-\gamma^2}(a-2m)^{\gamma^2-\gamma-1}\big[m^2\gamma(a\gamma-m(1+\gamma)\sin^2\theta+\nonumber\\
&&+a(a-2m)(a-m(1-\gamma))\big],\nonumber\\
D_{,r}(a,\theta)&=&2\sin^2\theta a^{\gamma}\left(a-m(1+\gamma)\right)(a-2m)^{-\gamma}.
\eeqs
It is now easy to check that these values for the radial derivatives of the metric functions do agree with the corresponding values computed using the exterior vacuum Zipoy-Vorhees geometry. In conclusion, the second fundamental form is also continue on the junction surface $r=a$ and this means that the interior solution (\ref{AZV}) will smoothly match the exterior Zipoy-Vorhees solution.

Now, let us turn our attention to the charged interior solutions described in the previous sections. As a general rule, after applying the Ehlers-Harrison transformations (electric or magnetic) then these metric functions will be transformed by multiplication by the factors $\Lambda^{\pm 2}$. Since $\Lambda$ is a smooth function of the coordinates $r$ and $\theta$ (as it is constructed algebraically from the metric component $A(r,\theta)$ or $D(r,\theta)$) then it should be clear that the above boundary conditions will be satisfied as well in the case of the charged interior solution. For example, in the electric case one has $\Lambda=1-E_0^2 A(r,\theta)^2$ and in the electrically charged solution the $tt$-component of the metric becomes $-\frac{A(r,\theta)^2}{\Lambda^2}$. This will be continuous and its first derivatives are continuous at the boundary $r=R$ as long as the seed function $A(r,\theta)^2$ is continuous and its first derivatives are continuous there. The remaining components will be multiplied by the factor $\Lambda^2$ and they will smoothly match the exterior geometry as well, since the corresponding metric components in the exterior geometry are multiplied by similar factors and they all agree on the matching hypersurface.

To see how this works in practice, let us consider first the electric solution generated by our method. The interior charged fluid solution is described by the new metric (\ref{electricm}). The normal radial direction for a surface of constant radius will be given by $\chi^a=(0, \Lambda B, 0, 0)$, where $\Lambda=1-E_0^2A^2$. The metric induced on a surface of constant radius will correspond to:
\beqs
h_{ab}dx^adx^b&=&-\frac{A^2}{\Lambda^2}dt^2+\Lambda^2\bigg[C^2d\theta^2+D^2d\varphi^2\bigg]
\eeqs

After one applies the same Harrison transformation on the external vacuum geometry one obtains a charged exterior geometry of the form:
\beqs
ds^2_{ext}&=&-\frac{\tilde{A}^2}{\tilde{\Lambda}^2}dt^2+\tilde{\Lambda}^2\bigg[\tilde{B}^2dr^2+\tilde{C}^2d\theta^2+\tilde{D}^2d\varphi^2\bigg].
\label{chargedv}
\eeqs

In the exterior charged geometry the metric induced on the same hypersurface $r=R$ is now
\beqs
\tilde{h}_{ab}dx^adx^b&=&-\frac{\tilde{A}^2}{\tilde{\Lambda}^2}dt^2+\tilde{\Lambda}^2\bigg[\tilde{C}^2d\theta^2+\tilde{D}^2d\varphi^2\bigg]
\eeqs
Here $\tilde{\Lambda}=1-E_0^2\tilde{A}^2$.

Continuity of the first fundamental form on the matching surface $r=R$ will require continuity of the metric components $\frac{A^2}{\Lambda^2}$, $\Lambda^2C^2$ and $\Lambda^2D^2$ on this surface. Since $A$ is continuous at this radius then also the factor $\Lambda$ is continuous there since $\Lambda=1-E_0^2A(r,\theta)$ and $\tilde{\Lambda}=1-E_0^2\tilde{A}(r,\theta)$. This means that when $A(R,\theta)=\tilde{A}(R,\theta)$ then also $\Lambda=\tilde{\Lambda}$ for $r=R$. 

Turning now our attention to the second fundamental form of the hypersurface $r=R$, let us notice that the components of the extrinsic curvature in the interior geometry take the simple form:
\beqs
K^t_t&=&\frac{\Lambda A_{,r}-A\Lambda_{,r}}{\Lambda^2 AB}, ~~~K^{\theta}_{\theta}=\frac{\Lambda C_{,r}-C\Lambda_{,r}}{\Lambda^2 CB},~~~K^{\varphi}_{\varphi}=\frac{\Lambda D_{,r}-D\Lambda_{,r}}{\Lambda^2 DB},
\eeqs
with similar tilded quantities for the components of the extrinsic curvature in the exterior geometry. Continuity of the second fundamental form requires then continuity of the functions $A(R,\theta)=\tilde{A}(R,\theta)$, $B(R,\theta)=\tilde{B}(R,\theta)$, $C(R,\theta)=\tilde{C}(R,\theta)$ and $D(R,\theta)=\tilde{D}(R,\theta)$ as well as the continuity of their partial derivatives in the radial direction $A_{,r}(R,\theta)=\tilde{A}_{,r}(R,\theta)$, $C_{,r}(R,\theta)=\tilde{C}_{,r}(R,\theta)$ and $D_{,r}(R,\theta)=\tilde{D}_{,r}(R,\theta)$. Note that the equality $\Lambda_{,r}=\tilde{\Lambda}_{,r}$ is assured by the continuity of the function $A$ and its first radial derivative $A_{,r}$.

As one can now easily notice, the smooth matching of the charged interior exterior geometries (\ref{electricm}) and (\ref{chargedv}) are assured by the smooth matching of the original interior geometry (\ref{initialm}) with the vacuum geometry  (\ref{initialmv}) as claimed. In particular, since the interior Zipoy-Vorhees solutions will smoothly match the exterior Zipoy-Vorhees solution on a surface of constant radius $r=a$, then by applying our charging technique on the interior as well as on the exterior geometry one should obtain a charged interior Zipoy-Vorhees solution which will smoothly match the exterior charged Zipoy-Vorhees geometry on the surface $r=a$. The same conclusions will apply to the magnetic case as well.

Another interesting issue is related to the electromagnetic potential since one has now one potential in the interior geometry as well as a potential in the exterior charged geometry. In the charged interior geometry one has an expression of the form (\ref{finalel}). In the exterior charged geometry the electromagnetic potential will have the expression:
\beqs
A_e&=&\frac{E_0\tilde{A}(r,\theta)^2}{\tilde{\Lambda}}.
\label{emv}
\eeqs
If the charged interior solution (\ref{electricm}) and the charged exterior (\ref{chargedv}) are smoothly matched, these conditions will assure the continuity of the electromagnetic potential as well as its radial derivatives across the surface $r=R$ and there will be no charged shell on the junction surface $r=R$.

One might wonder what happens if the interior geometry does not smoothly match the exterior geometry. In this case one has to use the Israel-Lanczos junction conditions to compute the thin-shell stress-energy tensor (\ref{Lanczos}) on the boundary surface. For example, based on the results in \cite{Stelea:2018cgm}, where the anisotropy induced by the magnetic field manifested itself in the superficial stress-energy tensor, we expect this to be the generic case, that is, the electrically charging/magnetizing effects manifest themselves in the boundary stress-energy tensor and their explicit forms should be computed on a case by case basis.

Finally, regarding the energy conditions one should mention that if they are satisfied in the initial seed solution then they should be generally preserved in the final charged solution as well. For example, consider a geometry with spherical symmetry for convenience. If the initial fluid seed satisfies say the Dominant Energy Condition (DEC) then:
\beqs
\rho^0-p_r^0\geq 0, ~~~ \rho^0-p_t^0\geq 0.
\eeqs
Consider now, for instance, the effect of the electrically charging transformation from Section $3$. The final energy density becomes $\rho=\frac{\rho^0}{\Lambda^2}+\rho_{el}$, while $p_r=\frac{p_r^0}{\Lambda^2}$ and $p_t=\frac{p_t^0}{\Lambda^2}$, where $\rho_{el}$ is given by (\ref{rhoel}) and it is manifestly positive. The net result is that DEC is satisfied by the electrically charged solution. A similar situation happens in the magnetic case as well.

\section{Conclusions}

In this work we presented two solution-generating techniques, which are direct generalizations of some of the Ehlers-Harrison transformations in the Ernst formalism, while adapted to work in presence of an anisotropic fluid source with axial symmetry. Based on these procedures we were able to construct the electrically charged and the magnetized solution for every static axially-symmetric geometry (\ref{initialm}), sourced by a anisotropic fluid described by a non-diagonal anisotropic stress-energy tensor (\ref{initialf}). While one might object that this is not usually the traditional way in finding solutions of the Einstein equations, one can also understand our solution-generating methods in the sense that for instance, if one uses the special metric ansatz as in (\ref{electricm}), an ansatz for the electric field (\ref{finalel}) and an ansatz for the fluid quantities (\ref{finalf}) then the problem of solving the Einstein-Maxwell-fluid equations is essentially reduced to find a solution of the Einstein-fluid equations for the metric (\ref{initialm}) and fluid quantities (\ref{initialf}). Conversely, from every fluid solution one can find its electrically charged version or the magnetized version.

As simple enough examples, we showed how to derive two new solutions describing the electrically charged version of the Bowers and Liang solution, as well as a magnetized version of an exact solution with axial symmetry presented in \cite{Herrera:2013hm}. This last solution describes an incompressible spheroid with homogeneous energy density and isotropic pressures, however its exterior geometry is unknown since it cannot belong to the Weyl-Papapetrou class. As such, its physical relevance is lacking at this point.

As the most important results of our solution-generating technique we also derived and analyzed two new solutions with axial symmetry that describe the electrically charged Zipoy-Vorhees interior solution as well as the magnetized Zipoy-Vorhees interior solution. Note that using our method one should be able to construct the electrically charged or the magnetized version of every static  axially symmetric fluid solution. As shown in Section $5$ if this solution can be smoothly matched to an exterior vacuum geometry in the Weyl class, then our transformations will produce the electrically charged or the magnetized versions of both the interior fluid solution as well as of the exterior geometries while still maintaining the smooth matching on the matching hypersurface. This method was previously used in \cite{Stelea:2018shx}, where based on the charging technique of interior fluid solutions with spherical symmetry we were able to obtain a new bound of the mass-to-radius ratio for electrically charged stars in GR. However, we stress again, our solution-generating technique can be successfully applied to more general interior solutions with axial symmetry as found for instance in \cite{Herrera:2013hm}, \cite{Hernandez-Pastora:2016ctg}.

Note that the matching is done here on a spheroid surface with constant radius $r=R$. In more general cases the matching can be done a surface $f(r,\theta)=0$ and the analysis performed in this paper should be generalized accordingly.

As avenues for further work, the magnetized solutions presented in our paper should be suitable to construct more realistic models of magnetars, by adding the slow-rotation in a perturbative way, along the lines of \cite{Hartle:1967he}, \cite{Benhar:2005gi}.  Another interesting extension of the present work would involve a study of the star’s anisotropy effect on the propagation of various fields in this background, on the lines of the study presented in \cite{Dariescu:2017ima}, \cite{Dariescu:2018dyy}, \cite{Dariescu:2010zz}.  Finally, it should be intersecting to study the extension of these results in spaces with higher dimensions using a solution generation procedure as in \cite{Stelea:2009ur}. 

Work on these matters is in progress and it will be presented elsewhere. 

\vspace{10pt}

\section*{Acknowledgements}

The authors would like to thank the anonymous referees whose remarks and suggestions helped improve this manuscript.

\end{document}